\documentclass[a4paper]{amsart}
\usepackage[dvips]{graphicx}
\usepackage{psfrag}

\newcommand{\Gr}{{\mathcal G}}

\newcommand{\V}{{\mathcal V}}
\newcommand{\C}{{\mathcal C}}

\newcommand{\CV}{{ C ( \V)}}
\newcommand{\CcV}{C_c (\V)}

\newcommand{\E}{{\mathcal E}}
\newcommand{\F}{{\mathcal F}}
\newcommand{\G}{{\mathcal G}}

\newcommand{\R}{\mathbb{R}}
\newcommand{\N}{\mathbb{N}}

\newtheorem{proposition}{Proposition}
\newtheorem{definition}[proposition]{Definition}
\newtheorem{theorem}[proposition]{Theorem}
\newtheorem{lemma}[proposition]{Lemma}

\title[Elliptic operators on graphs and eigenfunctions]{
Elliptic operators on planar graphs: Unique
continuation for eigenfunctions and nonpositive curvature}
\author[S.~Klassert, D.~Lenz, N.~Peyerimhoff, P.~Stollmann]{Steffen
Klassert, Daniel Lenz,\\ Norbert Peyerimhoff and Peter Stollmann}

\date{}

\begin{document}

\maketitle

\vspace{0.3cm}
\noindent
$^{1}$ Fakult\"{a}t f\"{u}r
Mathematik, Technische
Universit\"{a}t Chemnitz,
D-09107 Chemnitz, Germany;
E-mail:  S.Klassert@mathematik.tu-chemnitz.de,\\[0.1cm]

\noindent
$^{2}$ Fakult\"{a}t f\"{u}r
Mathematik, Technische
Universit\"{a}t Chemnitz,
D-09107 Chemnitz, Germany;
E-mail:  D.Lenz@mathematik.tu-chemnitz.de,\\[0.1cm]

\noindent
$^{3}$ Fakult\"{a}t f\"{u}r
Mathematik, Ruhr-Universit\"at Bochum,
D-44780 Bochum, Germany;
E-mail:  peyerim@math.ruhr-uni-bochum.de,\\[0.1cm]

\noindent
$^{4}$  Fakult\"{a}t f\"{u}r Mathematik, Technische Universit\"{a}t
Chemnitz,
D-09107 Chemnitz, Germany; E-mail:  P.Stollmann@mathematik.tu-chemnitz.de

%%%%%%%%%%%%%%%%%%%%%%%%%%%%%%%%%%%%%%%%%%%%%%%%%%%%%%%%%%%%
% ABSTRACT
%%%%%%%%%%%%%%%%%%%%%%%%%%%%%%%%%%%%%%%%%%%%%%%%%%%%%%%%%%%%

\begin{abstract} \noindent  
This paper is concerned with elliptic operators on plane
tessellations. We show that such an operator does not admit a
compactly supported eigenfunction, if the  combinatorial
curvature of the tessellation is nonpositive. Furthermore, we show
that the only geometrically finite, repetitive plane tessellations with
nonpositive curvature are the regular $(3,6), (4,4)$ and $(6,3)$
tilings.
\end{abstract} 

%%%%%%%%%%%%%%%%%%%%%%%%%%%%%%%%%%%%%%%%%%%%%%%%%%%%%%%%%%%%
% INTRODUCTION
%%%%%%%%%%%%%%%%%%%%%%%%%%%%%%%%%%%%%%%%%%%%%%%%%%%%%%%%%%%%

\section{Introduction} 
Since the work of Aronszajn \cite{Aro}, unique continuation for
elliptic operators on Riemannian manifolds and Schr\"odinger operators
has been investigated in
very many  papers of which we only mention \cite{Ag,ABG,H,JK}. It was
shown under rather general assumptions that a function $f$ on a
connected Riemannian manifold vanishes identically, whenever it
vanishes in one point to infinite order and satisfies
\begin{equation}\label{eigen}
H f =0
\end{equation}
where $H$ is an elliptic operator. In particular, $f$ satisfying
\eqref{eigen} must vanish identically if it vanishes on the non empty
complement of a a compact set.

\medskip

Despite all analogies between elliptic operators on graphs and those
on manifolds, unique continuation does not hold on graphs. It fact, it
is rather easy to give examples of elliptic operators on graphs with
compactly supported eigenfunctions. These examples have recently
attracted some attention since they play a role in the investigation
of the so called integrated density of states for random operators
\cite{DLMSchY,KLS,Ves}.

\smallskip

More precisely, in \cite{DLMSchY}, Dodziuk et al. study a certain
periodic Laplacian on graphs viz Laplacian on infinite graphs, which
are coverings of finite graphs by amenable groups. They show that the
eigenvalues of the whole graph operator are the union of the
eigenvalues of suitable restrictions to finite graphs. Moreover, they
obtain a characterization of the points of discontinuity of the
integrated density of states by existence of compactly supported
eigenfunctions.

Independently, existence of compactly supported eigenfunctions was
studied by three of the authors in \cite{KLS} for certain
aperiodically ordered graphs. This study does not only give examples
of compactly supported eigenfunctions, but again links their occurrence
to discontinuities of the integrated density of states.

More recently, there is related work of Veseli\'{c} on high random
graphs \cite{Ves}.

While a common framework to operators on these three classes is
still missing, the link between discontinuities of the integrated
density of states and the occurrence of compactly supported eigenfunctions is
by now well established.

\smallskip

The aim of this paper is to investigate  combinatorial
conditions on the graph which guarantee nonexistence of compactly
supported eigenfunctions.

In this context, the only result available so far is due to Delyon/Souillard
\cite{DS}. They show absence of compactly supported eigenfunctions for
random Schr\"odinger operators on the $d$-dimensional lattice and use
this to conclude continuity of the corresponding integrated density of
states.

Here, we will restrict our attention to plane tessellating graphs and
establish a connection between absence of compactly supported
eigenfunctions of an elliptic operator and the combinatorial curvature 
of the graph introduced in \cite{BP1}. Our main result uses
the geometric/combinatorial methods developped in \cite{BP1,BP2} and states
the following:

\smallskip

\textbf{Result 1} \textit{ If the curvature of the plane tessellation $\Gr$ is
nonpositive then no elliptic operator on $\Gr$ admits a compactly
supported eigenfunction. }

\medskip

Note that our result implies Delyon/Souillard's result in the particular
case of a two-dimensional lattice.

While the result apriori applies to general plane tessellating graphs,
it is of limited use when it comes to application to 
geometrically more rigid tessellations with certain repetitivity
properties (as are the ones encountered in \cite{KLS}). Namely, our
second result shows:

\smallskip

\textbf{Result 2} \textit{ If a geometrically finite, repetitive plane
tessellation has nonpositive curvature then it coincides with one of
the three regular combinatorial tessellations $(3,6)$, $(4,4)$ or
$(6,3)$. }

%%%%%%%%%%%%%%%%%%%%%%%%%%%%%%%%%%%%%%%%%%%%%%%%%%%%%%%%%%%%
% NOTATION AND RESULTS
%%%%%%%%%%%%%%%%%%%%%%%%%%%%%%%%%%%%%%%%%%%%%%%%%%%%%%%%%%%%

\section{Notation and results}

In this section we introduce the model which we consider and present
our results.

\medskip

A planar graph $\Gr = (\V,\E,\F)$ is a graph which is embedded in the
plane $\mathbb R^2$. Here, $\V$ denotes the set of vertices and $\E$
the set of edges of $\Gr$.  $\F$ denotes the family of connected
components of the complement of the image of $\Gr$ in $\mathbb
R^2$. The elements of $\F$ are called {\em faces} of $\Gr$.  We will
always assume that our graphs are compactly finite, i.e.  every point
in $\R^2$ has a neighbourhood which meets only finitely many faces.

We call a face $f\in\F$ a {\em polygon\/} if $\bar f$ is homeomorphic
to a closed disc and its boundary defines a simply closed path of
vertices. In this case, the boundary is called the {\em boundary
cycle\/} $\partial f$ of $f$.  The edges which belong to $\partial f$
are called the {\em sides} of $f$.  The number of edges of $f$ is
denoted by $E_{\partial f}$. If the polygon $f$ has $k$ sides
it is called a {\em $k$-gon}.  The number of edges emanating from a
vertex $v \in \V$ is called the {\em degree of $v$}, denoted by $|v|$.
Two vertices are called \textit{adjacent} if they are connected by
an edge. We write $v\sim w$ if $v$ and $w$ are adjacent vertices. 

\begin{definition}
A planar graph $\Gr$ in $\mathbb R^2$  is called {\em tessellating},
if the following conditions are satisfied:
\begin{enumerate}
\item[i)] Any edge is a side of precisely two different faces. 
\item[ii)] Any two faces are disjoint or have precisely either a vertex or
a side in common.
\item[iii)] Any  face $f \in \F$ is a polygon with finitely many sides.
\item[iv)] Every vertex has finite degree.
\end{enumerate}
\end{definition}

In order to present our results we have to introduce the 
corresponding notions of combinatorial curvature  and of 
elliptic operators.

 A \textit{corner} of a tessellating
graph is a pair $(v,f)\in\V \times \F$ so that $v \in \partial f$. The
set of all corners of $\Gr$ is denoted by $\C:= \C (\Gr)$.

\begin{definition}
Let $\Gr$ be a plane tessellation. Then, the 
function $\kappa: \C\to \mathbb R$ defined by 
\[
\kappa(v,f) := \frac{1}{|v|} + \frac{1}{E_{\partial f}}- \frac{1}{2}
\]
is called the the {\em curvature} (on the graph $\Gr$).  The graph
$\Gr$ is said to have nonpositive curvature if $\kappa (v,f)\leq 0$
for every $(v,f)\in \C$.
\end{definition}

\smallskip

Let $\CV$ be the vector space of all complex valued functions on
$\V$. Let $\CcV$ be the subspace of $\CV$ consisting of functions which vanish outside a finite set of vertices. 

A linear operator $L : \CV \longrightarrow \CV$ is called  {\em elliptic} 
if its matrix   $ a : \V \times
\V\longrightarrow \R$ satisfies   $a(v,w) \neq 0$ whenever $v \sim w$
and $a(v,w) = 0$ whenever $v \not\sim w$ and $v
\neq w$. Thus
\[
Lu(v) := \sum_{w \in \V} a(v,w)u(w) = a(v,v) u(v) + \sum_{w \sim v}
a(v,w) u(w)
\]
for every vertex $v$ in $\Gr$. A well studied example is the nearest
neighbour Laplacian where $a(v,v) = |v|$ for all $v\in \V$  and $a(v,w) = 1 $
whenever $v \sim w$.

As usual, a non vanishing function $u\in \CV$ is called an
eigenfunction of the elliptic operator $L$ to the eigenvalue $\lambda$  if $ L u - \lambda u = 0$, i.e. if
\[ \sum_{w \sim v} a(v,w)u(w) = (\lambda - a(v,v)) u(v)  
\quad \text{for all $v \in \V$}. \]

\medskip

%The aim of this article is to study occurrence of 
%compactly supported eigenfunctions of elliptic operators 
%on planar tessellations with non positive curvature. 
%Here, an eigenfunction $u$ of an elliptic operator
%is called compactly  supported if it vanishes in all but finitely many points. 

%It is proved in \cite{KLS} that there is an one to one 
%correspondence between the existence of eigenvalues with
%compactly supported eigenfunctions
%and the existence of jumps of the integrated density of states
%of the operator.
%These phenomena motivate to ask for a geometric criterion which
%guarantees or excludes the existence of compactly supported
%eigenfunctions. While we lack results for the first question, we present
%the following geometric criterion for the absence of compactly supported
%eigenfunctions in plane tessellations:

The precise version of our main result now reads as follows: 

\begin{theorem} \label{nolocaleigen}
Let $\G = (\V,\E,\F)$ be a plane tessellation with nonpositive
curvature and $L$ an elliptic operator on $\CV$. Then $L$  does not admit an
eigenfunction in $\CcV$.
\end{theorem}

Note that the above theorem takes only the combinatorial structure of
the tessellation into account and that the precise geometric shape of
the faces is of no importance. Our next result, however, is concerned
with geometrically more rigid tessellations $\Gr$. We assume that
$\Gr$ is build up by copies of finitely many fixed geometric tiles
$f_1,\dots,f_N \in \F$ and that every finite configuration of tiles in
$\Gr$ can be found repeatedly in any sufficiently large Euclidean
ball.

\begin{definition} \label{rep}
Let $\G = (\V,\E,\F)$ be a plane tessellation. Then $\G$ is called
{\em geometrically finite}, if there are finitely many faces $\{ f_1,
\dots, f_N \} \subset \F$, the {\em generators} of $\Gr$, such that
every $f \in \F$ is an isometric image of one of the generators, i.e., $f =
t + A f_j$, for a $j \in \{ 1,\dots,N \}$, $t \in \R^2$ and $A \in
SO(2)$. A geometrically finite tessellation $\G$ is called {\em
repetitive}, if for every finite set of faces $\{ g_1, g_2, \dots, g_k
\} \subset \F$ there is an $R > 0$ with the following property: In any
Euclidean ball $B_R(x) \subset \R^2$ there are $k$ faces 
$\hat g_1, \hat g_2, \dots, \hat g_k \in \F$ and $t \in \R$,
$A \in SO(2)$, such that $\hat g_j = t + A g_j$, for $j=1,\dots,k$.
\end{definition}

\begin{theorem} \label{quasicrystalpositive}
A geometrically finite, repetitive plane tessellation $\G$ 
of {\em nonpositive curvature} coincides with one of the three regular 
combinatorial tessellations $(3,6)$, $(4,4)$ or $(6,3)$.
\end{theorem}

The three regular tessellations are illustrated in \cite[Figure 1.2.1]{GS}.

%%%%%%%%%%%%%%%%%%%%%%%%%%%%%%%%%%%%%%%%%%%%%%%%%%%%%%%%%%%%
% GEOMETRY OF PLANE TESSELLATIONS
%%%%%%%%%%%%%%%%%%%%%%%%%%%%%%%%%%%%%%%%%%%%%%%%%%%%%%%%%%%%

\section{Geometry of plane tessellations}

In this section we discuss some geometric aspects of plane tessellations
following \cite{BP1,BP2}. Moreover, for non positively curved
plane tessellations we prove the impossibility of a certain
vertex labeling of the boundary cycle of distance balls.

\medskip

Let a plane tessellation $\Gr = (\V,\E,\F)$ be given. 

\smallskip

Two faces in $\F$ are called {\em neighbours}, if they have an edge in
common. A sequence $(f_0,f_1,\ldots,)$ of faces is a {\em (connected)
path}, if any two subsequent faces $f_j,f_{j+1}$ are neighbours. If
this sequence is finite i.e. of the form $(f_0,f_1,\ldots,f_n)$, the
length of this path is defined to be $n$.  The {\em (combinatorial)
distance\/} $d(f,g)$ of two faces $f$ and $g$ is the smallest number
$n$ for which there exists a path $(f_0,f_1,\ldots, f_n)$ with $f_0 =
f$ and $f_n = g$.

Similarly, for a finite set $P$ in $\F$ and $f\in \F$ arbitrary, we
define $d(f,P) :=\min\{d(f,g) : g\in P\}$. Thus, for each finite set
$P$ and any $k\in \N_0$ we can define the $k$-neigbourhoods of $P$  by
$$ B_k (P) := \{ f\in \F \mid d(f,P) \leq k\}.$$

A special example of much relevance in our considerations arises if
$P$ consists of only one element: For a given face $f_0$ and $k\in
\N_0$, we define the {\em distance balls}
\begin{equation}
B_k(f_0) = \{ f \in \F \mid d(f,f_0) \le k \}.
\end{equation}
 Accordingly, we define the {\em distance spheres} by
\begin{equation}
A_k(f_0) = \{ f \in \F \mid d(f,f_0) = k \}.
\end{equation}
Apparently, each distance ball arises from the previous distance ball
by adding a distance sphere.  

\smallskip

It turns out that this gives a very well behaved ``onion-like''
layered structure on the distance balls, provided  $\Gr$ has nonpositive
curvature.  This layered structure is the crucial ingredient in
the proofs of our main results. It was analyzed in detail in the works
\cite{BP1,BP2}. 

In order to discuss the relevant results we need some more definitions.

\smallskip

A finite subset $P$ of $\F$ is called a {\em polygon}, if $\bigcup_{f
\in P} \overline{f} \subset \R^2$ is homeomorphic to a closed
disc. Then,  $\partial P$  denotes the cycle of boundary vertices. If $P$
is a polygon and $v$ belongs to $\partial P$, we define the {\em inner
degree of $V$ with respect to $P$} by
$$|v|_P^i := \mbox{ number of faces of  $f \in P$ meeting in
$v$}$$
and the {\em exterior degree of $v$ with respect to $P$} by 
$$ |v|_P^e := |v| - |v|_P^i.$$

We introduce now a labeling of the boundary vertices of the polygon
$P$ with the letters $a$ and $b$. First, for any vertex $v$ of $\Gr$,
we define
\[ N(v) = \min \{ E_{\partial f} \mid f \in \F: v \in \partial f \}. \]
A vertex $v \in \partial P$ obtains label $a$ if $|v|^i_P = 1$, or if
both $N(v)=3$ and $|v|^i_P \leq 3$ hold. All other vertices of $\partial P$
obtain label $b$. Moreover, if $v$ is an $a$-vertex
satisfying $|v|^i_P = 1$ then we say that $v$ is of type $a^+$. One should
think of the labeling as a sort of book-keeping of convexity: $a^+$-vertices
of $\partial P$ are considered to be particularly convex and $b$-vertices 
to be particularly concave.  

\smallskip

The following proposition gives a simple fact on this labeling. 

\begin{proposition} \label{belabel}
Let $\Gr$ be a plane tessellation with nonpositive curvature.  Let $P$
be a polygon. If $v\in \partial P$ satisfies $|v|_P^e =1$, then no
edge starting in $v$ belongs to $\partial B_1 (P)$.  Moreover, in this
case $v$ has a $b$-label.
\end{proposition}
\begin{proof} 
The first statement is obvious from the definition of the exterior
degree. If $v$ would be an $a$-vertex then it would satisfy $|v| \le 4$ and
it would be adjacent to a triangle, in contradiction to nonpositive curvature. 
\end{proof}

We say $w = (v_0, v_1, v_2 \ldots, v_k)$ is a connected (vertex-)path
of length $|w|= k$ in $\Gr$ if all subsequent vertices of $w$ are
connected by an edge. A connected path $w \subset \partial P$ is
called {\em admissible\/} if for every vertex $v \in w$ with label $b$
its neighbours in $w$ carry the label $a^+$. We call the polygon $P$
{\em admissible} iff $\partial P$ is admissible. Thus, admissible
polygons have the property that particularly concave vertices are
compensated by particularly convex vertices in their neighbourhood.

\smallskip

It turns out that admissibility is preserved under taking
$k$-neigbourhoods if the curvature in nonpositive. More precisely, the
following is proven in Proposition 2.5 and Proposition 2.6 of
\cite{BP2}:

\begin{lemma} \label{dist1tubea}
Let $\Gr$ be a plane tessellation with nonpositive curvature.  Let $P$
be an admissible polygon. Then the set $B_1 (P)$ is an admissible
polygon. Moreover, for every face $f \in B_1 (P) - P$,
\begin{itemize}
\item[(a)] $\partial f \cap \partial P$ is a connected path of edges of 
length $\le 2$ and 
\item[(b)] $\partial f \cap \partial B_1 (P)$ is a connected path of
length $\geq 1$.
\end{itemize}
\end{lemma}

Apparently, every face $f\in \F$ is admissible. Thus, the lemma
immediately implies (see Corollary 2.7 in \cite{BP2} as well):

\begin{lemma}\label{BP27}
Let $\Gr = (\V,\E,\F)$ be a plane tessellation with nonpositive
curvature. Let $f \in \F$ be arbitrarily. Then, every distance ball
$B_k(f)$ is an admissible polygon and every face of the distance
sphere $A_k(f)$ contributes at least one edge to the boundary
$\partial B_k (f)$.
\end{lemma}

Now, the main result of \cite{BP2} is the following combinatorial
analogue of the Hadamard-Cartan theorem in differential geometry.
Note, that it can be seen as describing a nice layered structure of
$\Gr$.

\begin{theorem}\label{cl} Let $\Gr =(\V,\E,\F)$ be a plane tessellation. 
For a given face $f \in \F$, we define the cut locus $C(f) \subset \F$
of $f$ in the metric space $(\F,d)$ to be the set 
\[ C(f) := \{ g \in \F \mid d(f',f) \le d(g,f)\ \text{\rm for all neighbours
$f'$ of $g$} \}, \]
i.e., the set of all faces on which the distance function $d_f(g) = d(f,g)$
attains a local maximum. If $\Gr$ has nonpositive curvature, then $C(f) = 
\emptyset$ for all $f \in \F$.
\end{theorem}

This theorem has the following consequence: 

\begin{lemma}\label{BP27iii}
Let $\Gr = (\V,\E,\F)$ be a plane tessellation with nonpositive
curvature. Let $f \in \F$ be arbitrarily. Then the boundary $\partial
B_k (f)$ imposes a cyclic enumeration $(f_1,\ldots, f_n)$ of $A_{k+1}
(f)$ such that precisely subsequent faces intersect and each
intersection contains a vertex $v\in \partial B_k (f)$.
\end{lemma}
\begin{proof} This  follows from Theorem 3.2 of \cite{BP1}
and the fact that $\Gr$ has no cut locus by the previous theorem.
\end{proof}

We are heading towards the result on the vertex labeling mentioned in
the beginning of this section.

We will use the following lemma from \cite{BP2}. The lemma is
rather technical but very important for the proof of Theorem
\ref{nolocaleigen}.

\begin{lemma}[Lemma 2.8 in \cite{BP2}] \label{BP2-lemma28}  
Let $\Gr$ be a plane tessellation with nonpositive curvature and
$A_k(f) = \{f_1,f_2,\ldots,f_n \}$ be a distance sphere in $\G$ with
cyclic enumeration of its faces. Then there is at least one face $f_j
\in A_k(f)$ with one of the following properties: either $| \partial
f_j \cap \partial B_{k-1}(f) | = 1$ or $f_j$ does not share a common
edge with both $f_{j-1}$ and $f_{j+1}$ (mod $n$).
\end{lemma} 

The following proposition is a key step in our proof. It may be of
independent interest.

\begin{proposition}\label{label-seq}
Let $\G = (\V,\E,\F)$ be a plane tessellation with nonpositive
curvature and $B_k$ be a distance ball for a given
$f_0\in\F$ with the closed simple sequence of vertices
$v_0,v_1,v_2\ldots,v_{l-1},v_l = v_0$ describing $\partial B_k$.

If $l$ is even and every $b$-vertex on $\partial
B_k$ satisfies $|v|^e_{B_k} = 1$, then the closed simple label-sequence
of vertices describing $\partial B_k$ is not of the form $a^+$, $b$,
$a^+$, $b$, $a^+$, $b$, $\ldots$, $a^+$, $b$.
\end{proposition}

\begin{proof}
Note that $\partial B_0 = \partial f_0$ is of the form
$a^+,a^+,\dots,a^+$.  Let $k \ge 1$ and assume that the label sequence
of $B_k$ is of the form $a^+,b,\dots,a^+,b$. Then, for every $f \in
A_k := A_k(f_0)$, $\partial f \cap \partial B_k$ is a connected path
of length $\le 2$ (for otherwise there were at least two successive
vertices of label $a^+$).  By Lemma \ref{BP27iii}, the faces of $A_k$
can be enumerated consecutively such that only faces with subsequent
indices intersect. We call faces in $A_k$ with such a nonempty
intersection ``neighbours in $A_k$''.

\smallskip

\textit{Case 1:  There is a $b$-vertex $v$  of $\partial B_k$ satisfying
$|v|_{B_k}^i = 2$.}

As $|v|^e_{B_k} = 1$, we have $|v| = 3$. Thus, non positivity of the
curvature implies that all faces adjacent to $v$ have to be at least
$6$-gons.  Let $f \in B_k$ be a face with $v \in \overline{f}$. Note
that $f$ must belong to $A_k$, as $|v|=3$. Therefore, the edges of $f$
fall into three different types: those which belong to $\partial
B_{k-1}$, those which belong to $\partial B_k$ and those which are a
common edge with one of the two neighbours of $f$ in $A_k$.

As all distance balls are admissible, Lemma \ref{dist1tubea} (a) gives
$|\partial f \cap \partial B_{k-1}| \le 2$. Together with the fact $
|\partial f \cap \partial B_k| \le 2$ we conclude that $f$ has to be a
$6$-gon, sharing an edge with both its neighbours. Moreover, both
$b$-vertices of $\partial B_k$ belonging to $\overline{f}$ satisfy
$|v|_{B_k}^i = 2$.

By repetition of the above arguments this implies that, consecutively,
every face of $A_k$ has to be a $6$-gon, sharing an edge with both its
neighbours. This situation is not possible, by Lemma
\ref{BP2-lemma28}.

\smallskip

\textit{ Case 2: Every $b$-vertex of $\partial B_k$ satisfies
$|v|_{B_k}^i \ge 3$.} 

This means that none of the faces $f$ in $A_k$ shares a common edge
with any of its neighbours in $A_k$, and since $\partial f \cap
\partial B_{k-1}$ is a connected path of at most $2$ edges (see Lemma
\ref{dist1tubea} (a)), all faces $f \in A_k$ are at most $4$-gons.

\smallskip

\textit{ Case 2.1:   $A_k$ contains a $4$-gon $f$.}

The face $f$ contributes an edge to the boundary of $B_k$ and
therefore at least one of its edges carries an $a^+$ label. Thus,
$f$ contributes precisely two edges to the boundary of $B_k$ and
there exists a unique vertex $v$ of $\partial f$ which does not belong
to $\partial B_k$.

Assume that $v$ is an $a$-vertex with respect to the
labeling of $\partial B_{k-1}$.  Then $v$ is adjacent to a triangle
and we have $|v|^i_{B_{k-1}} \ge 5$, for curvature reasons, which
contradicts to label $a$. 

Hence, $v$ is an $b$-vertex with respect to
$B_{k-1}$. The neighbours of $v$ along $\partial B_{k-1}$ are then
$a^+$-vertices (since $B_{k-1}$ is admissible). Therefore, the two
$b$-vertices $v',v''$ with respect to $B_k$ which belong to $f$ are
also $a^+$-vertices with respect to $B_{k-1}$, i.e., we have
$|v'|^i_{B_k} = 3$. Thus we have $|v'| = 4$ because of $|v'|^e_{B_k} =
1$, and the neighbour $f'$ of $f$ in $A_k$ with $v' \in \partial f'$ cannot
be a triangle, for curvature reasons. 

This shows that all faces in
$A_k$ are $4$-gons and that we have, again, for $B_{k-1}$ the
situation that the vertices of $\partial B_{k-1}$ are labeled as
$a^+,b,a^+,b,\dots,a^+,b$ and all $b$-vertices satisfy
$|v|_{B_{k-1}}^e = 1$ and $|v|_{B_{k-1}}^i \ge 4$, for curvature reasons. 
Again, by curvature reasons, $A_{k-1}$ must
contain a $4$-gon, since there are vertices $v' \in \partial B_k \cap
\partial B_{k-1}$ with $|v'| = 4$. Thus, we may apply induction, and
conclude that $B_0 = \{ f_0 \}$ is a $4$-gon with label-sequence
$a^+,b,a^+,b$, which is a contradiction.

\textit{Case 2.2:  $A_k$ consists only of
triangles.}

The label sequence forces each triangle to contribute two edges to the
boundary of $B_k$. Let $f$ be a triangle in $A_k$. For curvature
reasons, the two $b$-vertices (with respect to $B_k$) of $f$ satisfy
$|v|^i_{B_{k-1}} \ge 3$ (note that $|v|_{B_k}^e=1$). This means that
the face $f'$ of $B_{k-1}$ which shares an edge with $f$ does not
share a common edge with both of its neighbours in $A_{k-1}$ and that
$|\partial f \cap \partial B_{k-1}| = 1$. Since
$\partial f' \cap \partial B_{k-2}$ is a connected path of at most $2$
edges, by Lemma \ref{dist1tubea} (a), $f'$ has to be, again, a triangle. 

Note that, by curvature reasons, the unique vertex of $f'$ which is not
also a vertex of $f$ has to be a $b$-vertex with respect to $B_{k-2}$.
Moreover, by admissibility of $B_{k-2}$, its label sequence is, again,
given by $a^+,b,$ $a^+,b,\dots,a^+,b$, and that all $b$-vertices of
$\partial B_{k-2}$ satisfy $| v |_{B_{k-2}}^e = 0$. Now, $B_{k-2}$
satisfies, again, the conditions of the proposition. Since neither
Case 1 nor Case 2.1 can be given, we conclude that, again, Case 2.2
is given for $A_{k-2}$, namely, $A_{k-2}$ consists only of triangles.  
We can repeat the same arguments inductively.

In the case that $k$ was even, induction leads to $B_0 = \{ f_0 \}$ being
a single triangle. Its three vertices would have to be labeled as
$a^+,b,\dots,a^+,b$, which is not possible for parity reasons. In the case
that $k$ was odd, we end up with $B_1$ consisting of the center face $f_0$ 
and triangles attached to each of the edges of $f_0$. The property 
$|v|_{B_1}^e=1$ of each $b$-vertex of $\partial B_1$ then implies
$|v| = 4$ which yields a contradiction to nonpositive curvature.  
\end{proof}

%%%%%%%%%%%%%%%%%%%%%%%%%%%%%%%%%%%%%%%%%%%%%%%%%%%%%%%%%%%%
% PROOFS OF THEOREM NOLOCALEIGEN AND QUASICRYSTALPOSITIVE
%%%%%%%%%%%%%%%%%%%%%%%%%%%%%%%%%%%%%%%%%%%%%%%%%%%%%%%%%%%%

\section{Proof of Theorem \ref{nolocaleigen}}

Let $\G =(\V,\E,\F)$ be a plane tessellation with nonpositive curvature. 
%Our aim is to prove $u \equiv 0$. 
We choose $f_0 \in \F$ and define $B_k := B_k (f_0)$ and $A_k = B_k
\backslash B_{k-1}$. Note that $B_0 = \{ f_0 \}$.  Let $u\in \CcV$   be a compactly supported eigenfunction.

%As in \cite{BP2}, we refer to $\partial B_k$ as a set of vertices. 
By Lemma \ref{BP27}, $B_k$ is a polygon for every $k\in \N_0$, i.e.,
the boundary of $B_k$ defines a simple closed path of vertices.  In
particular, the boundary is a Jordan curve and thus divides the plane
in an interior and an exterior part. Therefore, we can define the set
of vertices $\V_k$ by

$$ \V_k :=\{ \mbox{vertices on $\partial B_k$}\} \cup \{\mbox{vertices
outside of $B_k$} \}.$$

\smallskip

Theorem \ref{nolocaleigen} follows if we prove the following  two steps: 

\begin{enumerate}
\item[i)] There exists a $n_0\in \N$ such that $u$ vanishes on $\V_{n_0}$. 
\item[ii)] If $u$ vanishes on $\V_{k}$ for some $k\in \N$, then $u$
vanishes on $\V_{k-1} $ as well.
\end{enumerate}

Here, i) is immediate from the local finiteness of the graph. Thus,
the main point is to prove ii).

\medskip

To prove ii), we need the following lemma.

\begin{lemma}\label{lemma2}
Let $u\in \CcV$ be an eigenfunction of the 
elliptic operator $L$ on $\CV$  with 
$u \vert_{\V_{k+1}} \equiv 0$. Then, $u(v) =0$ for  all vertices 
$v \in \partial B_k$ with $|v|^e_{B_k} > 1$. 
\end{lemma}

\begin{proof} Let $v \in \partial B_k$. If $|v|^e_{B_k} > 2$ then,
obviously, $v$ is also a vertex of $\partial B_{k+1}$ and we have
$u(v) = 0$. Thus, we only have to consider the case $|v|^e_{B_k} = 2$:

Choose $f \in A_k$ with $v \in \partial f$. Let $f' \in A_{k+1}$ be a
neighbour of $f$. $|v|^e_{B_k} = 2$ implies that $f'$ shares a common
edge $e$, emanating from $v$, with a neighbour $f''$ in
$A_{k+1}$. Assume that $e$ connects the vertices $v$ and $v'$. Since
two faces have at most one edge in common, we conclude that $v'$
belongs to $\partial B_{k+1}$. Now, $v'$ cannot belong to $\partial
B_k$, for otherwise $f'$ would be in the cut locus $C(f_0)$ (which is
empty by Theorem \ref{cl}). Now, we inspect the other vertices
adjacent to $v'$.

\smallskip

\textit{Case 1: All vertices adjacent to $v'$ and different to $v$
belong to $\V_{k+1}$.}

Then, none of these vertices belongs to $\partial B_k$. Thus, we
conclude from the fact $u(v') = 0$ that
\[ 0 = (\lambda - a(v',v'))\, u(v') = a(v',v)\, u(v) + 
\underbrace{\sum_{v'' \sim v', v'' \neq v} a(v',v'')\, u(v'')}_{=0}, 
\] 
and thus $u(v) = 0$ as $a(v',v) \neq 0$ by ellipticity.  

\smallskip

\textit{Case 2: There is a vertex $w \neq v$ adjacent to $v'$ not belonging
to $\V_{k+1}$.}

%As $w$ is adjacent to $v'$, it must belong to $\partial B_k$. 
%If there are also vertices in $\partial B_k$ adjacent
%to $v'$ which are not in $\partial B_{k+1}$ and different 
As $w$ is different from $v$, we infer $|v'|^i_{B_{k+1}}>2$. Thus,
there is at least one face $\hat f \in \{ f', f'' \} \subset A_{k+1}$
which satisfies $v' \in \partial \hat f$ and $\partial \hat f\cap
\partial B_{k+1} =v'$. Therefore, the boundary cycle $\partial
B_{k+1}$ does not share an edge with the face $\hat f \in A_{k+1}$, and
we have $\hat f \in C(f_0)$, contradicting to Theorem \ref{cl}.
%Therefore if $v'$ is adjacent to vertices in $\partial B_k$
%different from $v$, they belong also to $\partial B_{k+1}$.
%The same reasoning as above implies also $u(v) =0$ in this case. 
\end{proof}

In the following we let
\begin{equation} \label{vertexseq}
v_0, v_1, v_2, \dots, v_{l-1}, v_l = v_0
\end{equation}
be the closed simple sequence of vertices describing $\partial
B_k$.

\begin{lemma}\label{uvanish}
Let $u\in \CcV $ be a compactly supported eigenfunction of an 
elliptic operator on $L$ with $u \vert_{\V_{k+1}} \equiv 0$.
If $u$ vanishes for two subsequent vertices of the
sequence \eqref{vertexseq}, then $u$ vanishes on all of $\partial
B_k$.
\end{lemma}

\begin{proof}
Assume that $u(v_j) = u(v_{j+1}) = 0$. The vertex $v_{j+2}$ satisfies
either $| v_{j+2} |_{B_k}^e > 1$, in which case we have $u(v_{j+2}) =
0$, by Lemma \ref{lemma2}, and we can continue by considering the
subsequent vertices $v_{j+1}, v_{j+2}$. Otherwise, we have $| v_{j+2}
|_{B_k}^e = 1$, in which case $v_{j+2}$ carries label $b$ with respect
to the polygon $B_k$, for curvature reasons. 

Since the distance ball
$B_k$ is admissible, it follows that $v_{j+1}$ is an
$a^+$-vertex, i.e., we have $| v_{j+1} |_{B_k}^i = 1$. This implies
that $v_j$ and $v_{j+2}$ are the only vertices in $\overline{B_k}$
which are adjacent to $v_{j+1}$.  Thus, all other vertices adjacent to $v_{j+1}$ belong to $\V_{k+1}$ and $u$ vanishes on them. 

Defining  $\V_{k+1}' := \V_{k+1} \backslash 
\{ v_j, v_{j+2} \}$, we can therefore calculate
\begin{eqnarray*}
0 &=& (\lambda- a(v_{j+1},v_{j+1}))\, u(v_{j+1}) \\ 
  &=& a(v_{j+1},v_j)\,
  \underbrace{u(v_j)}_{=0} +\, a(v_{j+1},v_{j+2})\, u(v_{j+2}) +
  \underbrace{\sum_{v \sim v_{j+1}, v \in \V_{k+1}'}
  a(v_{j+1},v)\, u(v)}_{= 0}\\
&=& a(v_{j+1}, v_{j+2}) u (v_{j+1}).
\end{eqnarray*}
Thus, by ellipticity, we conclude  $u(v_{j+2}) = 0$. 

Hence, we can also continue with the subsequent
vertices $v_{j+1}, v_{j+2}$ in this case. The lemma follows now by iteration.
\end{proof}

\begin{proof}[Proof of Theorem \ref{nolocaleigen}]
We follow our strategy and show that $u|_{\V_{k+1}}\equiv 0$ yields
$u|_{\partial B_k}\equiv 0 $.  

\smallskip

By Lemma \ref{lemma2}, $u$ vanishes on all vertices with $|v|_{B_k}^e
> 1$. If $v$ is an edge which does not satisfy $|v|_{B_k}^e > 1$, it
must satisfy $|v|_{B_k}^e =1$.  Then $v$ carries an $b$-label by
Proposition \ref{belabel}. As $B_k$ is admissible it is then
enclosed by two $a^+$-vertices $v_{j-1}$ and $v_{j+1}\, ({\rm mod}\, l)$. By
Proposition \ref{belabel} again, these vertices have exterior degree
at least $2$ and then, by Lemma \ref{lemma2}, $u$ vanishes on them. 

These considerations show that $u$ vanishes at least for every second
vertex of $\partial B_k$.
%In order to show this, we make use of
%the properties of distance balls.
%Let 
%\begin{equation} \label{vertexseq}
%v_0, v_1, v_2, \dots, v_{l-1}, v_l = v_0
%\end{equation}
%be the closed simple sequence of vertices describing $\partial
%B_k$. 
%Lemma \ref{dist1tubea} states that, for every face $f \in A_k$, the
%intersection $\partial f \cap \partial B_k$ is a connected path of at
%most $2$ edges. This implies that every vertex $v_j \in \partial B_k$
%with $|v_j|_{B_k}^e=1$ is enclosed by two vertices $v_{j-1}, v_{j+1}$
%(modulo $l$) with $|v|_{B_k}^e > 1$. So we know already that, by Lemma
%\ref{lemma2}, $u$ vanishes on all vertices with $|v|_{B_k}^e > 1$ and,
%therefore, $u$ vanishes at least for every second vertex of $\partial
%B_k$.  
Now, by Lemma \ref{uvanish}, there remains only one case
for $u \vert_{\partial B_k} \not\equiv 0$: $l$ in
\eqref{vertexseq} is even and, by the admissibility of distance balls,
the corresponding label-sequence is $a^+,b,a^+,b,a^+,b,\dots, a^+,b$,
where every $a^+$-vertex satisfies $|v|_{B_k}^i = 1$ (by definition)
and every $b$-vertex satisfies $|v|_{B_k}^e = 1$. But this case is impossible
by Proposition \ref{label-seq}.
\end{proof}

\section{Proof of Theorem \ref{quasicrystalpositive}}

The basic idea in the proof of Theorem \ref{quasicrystalpositive} is
that the existence of a face with a negatively curved corner together
with repetitivity implies exponential growth of the number of faces in
a combinatorial distance ball. On the other hand, geometrical
finiteness implies that combinatorial and Euclidean balls are
comparable and that the number of faces inside a Euclidean ball can
only grow quadratically with the radius. Obviously, both growth
properties are contradictory and $\Gr$ is forced to have zero curvature.
  
Assume that $\Gr = (\V,\E,\F)$ is geometrically finite (with
generators $f_1, \dots, f_N \in \F$) and repetitive. Nonpositive
curvature implies that we have no cut-locus.  We prove that 
$\kappa(v,f) = 0$ for all corners. Assume, there is $(v,f) \in \C$
with $\kappa(v,f) < 0$. By repetitivity, there is a constant $C > 0$
and a radius $R > 0$, such that every Euclidean ball $B_R(x) \subset \R^2$
contains a face $f$ with
\[ \chi(f) := \sum_{v \in \bar f} \kappa(v,f) \le -C. \]
We choose $f_0 \in \F$ arbitrarily and denote $B_k(f_0)$, shortly, by
$B_k$. By geometrical finiteness, there are constants $0 < d < D$ and
a point $x_0 \in \R^2$ such that we have, for large enough $k$:
\[ B_{k d}(x_0) \subset \bigcup_{f \in B_k} \bar f \subset B_{k D}(x_0). \]
By volume comparison, we immediately obtain the following two facts:
\begin{enumerate}
\item[i)] There is a constant $c_1 > 0$ such that, for $k$ large enough,
$B_{k d}(x_0)$ contains at least $c_1 k^2$ disjoint Euclidean balls of radius
$R$. 
\item[ii)] There is a constant $c_2 > 0$ such that, for $k$ large enough,
$B_{k D}(x_0)$ contains at most $c_2 k^2$ faces of $\F$.
\end{enumerate}
Therefore, the mean Euler-characteristic $\overline{\chi}(B_k) :=
(\sum_{f \in B_k} \chi(f)) / | B_k |$ of distance balls $B_k$
satisfies, for all $k$ large enough,
\[
\overline{\chi}(B_k) = 
\frac{1}{| B_k |}\sum_{f \in B_k} \chi(f) \le - \frac{c_1}{c_2}C,
\]
where $|B_k|$ denotes the number of faces in $B_k$. By the remark on
page 156 of \cite{BP1}, this implies that $| B_k(f_0) |$ grows
exponentially in $k$, which contradicts ii).

Consequently, the plane tessellation has zero curvature in all
corners, and this immediately yields for each corner $(v,f) \in \C$:
$(|v|,E_{\partial f}) \in \{ (3,6), (4,4), (6,3) \}$. Finally, face to
face extension forces $\Gr$ to be a regular tiling of type $(3,6),
(4,4)$ or $(6,3)$.

\section{Further Remarks}
In the previous sections we have undertaken some first steps into
investigating the geometric situation leading to compactly supported
eigenfunctions. We could show that their existence is connected to
curvature properties of the underlying graph in the two dimensional
situation. This raises various questions:

\begin{itemize}
\item Do similar results hold in arbitrary dimension?

\item What are sufficient condition for existence of compactly supported eigenfunctions?

\item Can one develop a general framework of random operators covering the connection between compactly supported eigenfunctions and the discontinuities of the integrated density of states?

\end{itemize}

We plan to attack these questions in the future.

\end{document}